\documentclass[preprint,12pt]{elsarticle}

\usepackage{amssymb}
\usepackage{amsmath} 
\usepackage{amsthm}
\usepackage{graphics,xcolor} 
\usepackage{algorithm}
\usepackage{algorithmic}
\usepackage{epsfig}
\usepackage{epstopdf}
\usepackage{multirow}

\journal{Physica A: Statistical Mechanics and its Applications}

\begin{document}

\begin{frontmatter}



\title{Centrality Measures for Networks with Community Structure}


\author{Naveen Gupta}
\address{Computer Science and Engineering Discipline\\
PDPM Indian Institute of Information Technology,\\
Design \& Manufacturing Jabalpur,\\
Jabalpur, India\\ email: {Corresponding author: naveengupta4790@gmail.com}}
\author{Anurag Singh}
\address{Department of Computer Science and Engineering,\\
National Institute of Technology,\\
Delhi, India\\}
\author{Hocine Cherifi}
\address{University of Burgundy, LE2I UMR CNRS 6306, \\Dijon, France}

\begin{abstract}
Understanding the network structure, and finding out the influential nodes is a challenging issue in the large networks. Identifying the most influential nodes in the network can be useful in many applications like immunization of nodes in case of epidemic spreading, during intentional attacks on complex networks. A lot of research is done to devise centrality measures which could efficiently identify the most influential nodes in the network. There are two major approaches to the problem: On one hand, deterministic strategies that exploit knowledge about the overall network topology in order to find the influential nodes, while on the other end, random strategies are completely agnostic about the network structure. Centrality measures that can deal with a limited knowledge of the network structure are required. Indeed, in practice, information about the global structure of the overall network is rarely available or hard to acquire. 
Even if available, the structure of the network might be too large that it is too much computationally expensive to calculate global centrality measures. To that end, a centrality measure is proposed that requires information only at the community level to identify the influential nodes in the network. Indeed, most of the real-world networks exhibit a community structure that can be exploited efficiently to discover the influential nodes. We performed a comparative evaluation of prominent global deterministic strategies together with stochastic strategies with an available and the proposed deterministic community-based strategy. Effectiveness of the proposed method is evaluated by performing experiments on synthetic and real-world networks with community structure in the case of immunization of nodes for epidemic control.

\end{abstract}


\end{frontmatter}


\section{Introduction} \label{s1}
Outbreak of infectious diseases is a serious threat to the lives of people and it also brings serious economic loss for the victim countries. So, it is very important to discover the propagation rules in the social groups in order to prevent the epidemics or at least to control the epidemic spreading. Vaccination allows to protect people and prevent them to transmit disease among their contacts. As mass vaccination is not always feasible, due to limited vaccination resources, targeted immunization strategies are of prime interest for public health. Impact of the contact network topology on disease transmission is a hot topic in the large networks study. A lot of work is done towards this direction by various researchers \cite{barthelemyvelocity,EpdInt,gong2013efficient,halloran2008modeling,pastorepidemic, singh2012rumour,Anuappb,anuncc}.

Network organization can be characterized at different scales ranging from the microscopic to the macroscopic level. At the microscopic level, one concentrates on the differences between individuals in order to identify the most influential ones. At this level, statistical measures are used to summarize some of the overall network features. The centrality of a node is a very important feature. Since the  nodes  with high centrality values can  propagate information to the whole network more easily than the rest of the nodes with low centrality measures, they are the ones to be targeted for immunization. Unfortunately, there is no general consensus on the definition of centrality and many measures are proposed. Degree centrality and betweenness centrality are the most influential methods \cite{Betcent1, degcent1}. They are studied extensively to understand their effect on epidemic dynamics. The degree centrality of a node is its number of links to other nodes, while the betweenness centrality measures its contribution to the shortest paths between every pair of nodes in the network. As most real world-networks follow a power law distribution, it is very efficient to immunize the hubs (i.e., nodes having high degrees) preferentially. However, many real world networks are more structured than merely having heterogeneous degree distributions.

Sometimes, there is no information available about the global structure of the real-world networks. Hence, centrality measures based on locally available network information are of prime interest. Such strategies rely only on local information around the selected nodes. The most basic strategy is random immunization where target nodes are picked at random regardless of the network topology. In acquaintance immunization, the selected node is chosen randomly among the neighbors of a randomly picked node. As randomly selected acquaintances have more connections than randomly selected nodes, this method targets highly connected nodes \cite{cohen}. It can be viewed as a local approximation of a global degree based strategy.

The mesoscopic level of organization is mainly concerned by properties shared by sets of nodes called communities. A community-structured network is organized around different communities such that the number of links connecting the nodes in the same community is relatively large as compared to the number of links connecting nodes in different communities. While the effect on epidemic dynamics of degree distribution is studied extensively, the community structure of networks has attracted little attention. Its investigation has recently started, despite the fact that many real-world networks show significant community structure. Results reported in the literature confirm that the knowledge of the network degree distribution is not sufficient to predict epidemic dynamics. Indeed, fundamentally different behaviors are observed due to variations of the community structure in the networks with the same degree distribution. Therefore, the design of immunization strategies needs to take into account the community structure. At this level, the focus is on structural hubs (i.e., the nodes connecting the largest number of different communities). Indeed, these nodes act as bridges facilitating the propagation of the epidemic from one dense community to another \cite{salathe, bridge}. The diffusion over the network is important to study for the epidemic spreading. Samukhin \textit{et al.} have studied the Laplacian operator of an uncorrelated random network and diffusion processes on the networks. They have proposed a strict approach for diffusion process \cite{r1}. An exact closed set of integral equations are derived. which provide the averages of the Laplacian operator's resolvent, it initiated to describe the propagation of a random walks on the network. Mitrovic \textit{et al.} have studied eigenvalue spectra and the diffusion process with random walk in a various class of networks  at mesoscopic level \cite{r2}. Arenas \textit{et al.} have studied the relationship among the topological scale and dynamic time scale in the networks based on the dynamics closer to synchronization of coupled oscillators. In their study they have considered the communities of nodes organized in a hierarchical manner at different time \cite{r3}.

Immunization strategies in community-structured populations are previously investigated. Masuda \cite{masuda} proposed an immunization strategy based on the eigenvector centrality for community-structured networks. In this paper, module based strategy technique is used for measuring the contribution of the each node to the weighted community network in order to preferentially immunize nodes that bridge important communities. The number of links they share gives the weight of a link between any two communities. The influence of a node is related to the importance of the community together with its connectivity to the other communities. The effectiveness of the method is compared to alternative centrality measures (degree, betweenness, eigenvector) in the real and synthetic networks with modular structure. Three community detection algorithms are used to uncover the non-overlapping communities. The generated networks used to simulate real-world networks are made of equal-sized communities with heterogeneous degree distributions.

Salath\'{e} \textit{et al.} \cite{salathe} proposed an immunization strategy and targeted at nodes that connect to multiple communities, it was called community bridge finder (CBF). CBF is a local algorithm based on random walk. Starting from a node chosen randomly, it follows a random path until it reaches a node that does not connect to more than one of the previously visited nodes. They have used standard epidemic spreading model, susceptible-infected-resistant (SIR) \cite{sir} for finding the results on both real-world and synthetic networks. The authors compared three global centrality measures: degree, betweenness and random walk, and two local strategies: acquaintance and CBF. It is shown that the random walk centrality is the most successful strategy. Furthermore CBF outperforms the acquaintance method. In this work, the communities on the empirical network are defined functionally, and the synthetic data are generated using the Watts and  Strogatz model \cite{sw}.\\

However, there are some major issues associated with these contributions to the problem. First of all, the community structure of real-world networks is often not known, and one must rely on a community detection algorithm in order to uncover it. Therefore, it is therefore difficult to measure the influence of this operation on the results. Indeed, the classes of techniques available to detect communities are both numerous and diverse, and until now, there is no consensus on either a community definition or a universal detection method. Even, when a ground truth based on functional features is available, the topology of the selected networks can hardly be diverse enough to represent all types of systems encountered in various applications. To overcome these limitations, artificial networks can be used. Indeed, a random model allows generating as many networks as desired, while controlling some of their topological properties. However, this strategy raises another issue: the realism of the obtained networks, which should mimic closely real-world networks in order to get relevant test results. It is worth mentioning that some attempts are made to develop a theoretical framework for understanding the effects of the modular structure in networks. Unfortunately, models used in previous work are oversimplified. Simple cases in which modules are homogeneous and of equal size are considered while in real-world networks, modules are found to be heterogeneous in various aspects. For example, it is well established that networks with a community structure are characterized by a power-law distributed community size \cite{palla, guimera, vespi, clauset}.

Our investigations, address the above limitations. In order to understand how community structure affects epidemic dynamics we perform an extensive investigation using a realistic generative model \cite{LFR} with controlled topological properties and a representative set of immunization strategies. The main goal of this study is to give a clear answer on which immunization strategy should be preferred in order to control epidemics in networks with community structure. Second, the impact of the community structure is analyzed  using the LFR benchmark algorithm \cite{LFR}. We also perform the experiments using the proposed method on some real-world networks to further validate results on synthetic benchmark. An extensive experimental evaluation is conducted in order to investigate their efficiency as compared to global centrality measures, random-walk based stochastic measures as well as another community-based strategy. The major contributions of this paper are twofold. First,  a new local centrality measure based on the network community structure  is introduced. Second, an extensive experimental evaluation is conducted in order to investigate its efficiency as compared to global centrality measures, random-walk based stochastic measures as well as an alternative community-based strategy. The impact of the community structure is analyzed using both controlled synthetic benchmark produced by the LFR algorithm \cite{LFR} and  real-world networks. The  SIR epidemic spreading model \cite{morsat} is considered in order to test  the immunization strategies.

The rest of the paper is organized as follows. In section \ref{s2}, various centrality measures and immunization strategies are presented. Classic epidemic spreading model is discussed to understand the diffusion over the heterogeneous networks. In  section \ref{s4}, the benchmark model is introduced and its main properties are recalled.  Section \ref{s5} is devoted to the experimental results. The observations and findings are described in section \ref{s6}.
 
\section{Immunization Strategies} \label{s2}
Targeted immunization strategies can be divided into three categories based on their requirement about the knowledge of the network topology. Global strategies exploit the knowledge of the full network structure in order to find the influential nodes while Community-based strategies are able to work with a limited amount of information. The third category of immunization strategies are stochastic strategies which rely only on local information around randomly selected nodes. These three types of strategies are described below in detail:
 
\subsection{\textbf{Global Strategies:}} These immunization strategies are based on an ordering of the nodes of the whole network according to an influence measure. Nodes are then targeted (removed) in the decreasing order of their rank. The influence of a node is computed according to some centrality measure. In this study, two prominent global centrality measures (degree and betweenness) are considered in order to compare with the proposed strategies.

\textbf{1. \textit{Degree Centrality}:} Degree centrality denotes the number of immediate neighbors of a node, i.e. which are only one edge away from the node. It is simple but very coarse. It can be interpreted as the number of walk of length one starting at the considered node. It measures the local influence of a node. It has many ties and fails to take into account the influence weight of even the immediate neighbors (Algorithm \ref{a1}). Even if it is a local measure, the immunization strategy is global because it needs to rank all the nodes of the network according to their degree. 

\begin{algorithm}
   \caption{To calculate Degree Centrality}    \label{a1}    
    \textbf{Input} {Graph $G(V, E)$}\\
    \textbf{Output} {A map/dictionary with (\textit{node, centrality value}) pairs}
    
    \begin{algorithmic}    
    
    \STATE  Initialize an empty map, $M$ with $(node, value)$ pair.\\
    \FOR {each node $u$ in set $V$:}
    
    \STATE  Calculate the number, $n$ of adjacent nodes or neighbors of $u$.
    \STATE  Add the pair ($u$, $n$) to the map $M$.
     \ENDFOR
    \STATE  Return the map $M$.
   
\end{algorithmic}
\end{algorithm}

\textbf{2. \textit{Betweenness Centrality}:} Betweenness centrality defines the influence of a node based on the number of shortest paths between every pair of nodes that passes through that node. It basically tries to identify the influence of a node in terms of information flow through the network (Algorithm \ref{a2}). In this strategy, the nodes are target based on their overall betweenness centrality. The computation of betweenness has high time complexity.

\begin{algorithm}
 \caption{To calculate Betweenness Centrality} \label{a2}
   \textbf{ Input} {Graph $G(V, E)$}\\
    \textbf{Output} {A map/dictionary with (node, centrality value) pairs}
    
    \begin{algorithmic}    
    
    \STATE Initialize an empty map, $M$ with (node, value) pair format.
    \STATE Calculate the number ($n$) of shortest paths between every pair of nodes ($a$, $b$) in the graph.
    \STATE Store the result in step 2 in a map, $M$ with pairs ( ($a$, $b$), $n$ ).
     \FOR{each node $u$ in set $V$:}
    
  \STATE  Let $BC(u)$ be the betweenness centrality of node $u$, and initialize it to zero.
    \FOR {each node pair ($a$, $b$) in the map $M$}
    
    \STATE Count the number ($n_1$) of shortest paths between the nodes $a$ and $b$, of which node $u$ is also a part of.
    \STATE $BC(u)$ = $BC(u)$ + $n_1$/$n$
    
    \ENDFOR
    
    \ENDFOR
    \STATE Return the map $M$.
   
\end{algorithmic}
\end{algorithm}

\subsection{\textbf{Stochastic Strategies:}} Stochastic strategies do not need any information about the global structure of the network. They are able to work with a very limited amount of local information about a node. They basically try to find the influential nodes based on random walks. Such algorithms, agnostic about the full network structure are necessary in some situations. For example, the full structure of the contact network relevant to the spread of a disease is generally not known.

\textbf{1. \textit{Random Acquaintance:}} Random Acquaintance immunization strategy is proposed by Cohen \textit{et al.} \cite{cohen}. It works as follows: pick a random node $v_0$, and then pick its acquaintance or neighbor, $v_1$ at random. Immunize the nodes which are picked as acquaintances at least n times. An acquaintance is immunized immediately in the case of $n=1$. This strategy identifies highly connected individuals without any information about the global structure of the network.

\textbf{2. \textit{Community Bridge Finder (CBF):}} The CBF algorithm, proposed by Salathe \textit{et al.} \cite{salathe}, is a random walk based algorithm, aimed at identifying nodes connected to multiple communities. The algorithm begins with selecting a random node as the starting node. Then a random path is followed until a node is found that is not connected to more than one of the previously visited nodes on the random walk. It is based on the idea that the first node not connecting back to already visited nodes of the current random walk is more likely to belong to a different community. This strategy is completely agnostic about the network structure.\\

\subsection{\textbf{Community-based Strategies: }}These strategies do not require any information about the global structure of the network. They only require information at the community level. In the case where the community structure is unknown, it can be uncovered with local community detection algorithms \cite{chen, martelot}. In a network with community structure, the degree of a node can be split into two contributions: the intra-community links connecting it to nodes in its community and  the inter-community links connecting it to nodes outside its community. Strength of the community structure of a network depends upon inter and intra-community links. A network is said to have a strong community structure if a small fraction of total links in the networks lies between the communities. It is said to have well defined communities. On the contrary, if a large fraction of total links lies between the communities, then the network does not contain well-defined communities and it is said to have a weak community structure.

The topology of a network can be fully specified by its adjacency matrix $A$. In the case of an undirected, unweighted network, $A(i,j)$ is equal to 1 if nodes $i$ and $j$ are directly connected to each other otherwise it is equal to zero. Considering a community $C$ of a network, the total degree of a node $i$ can be split in to two parts: 
\begin{equation} 
k_i(C) = k_i^{in}(C) + k_i^{out}(C).
\end{equation}
The degree $k_i$ of a node $i$ is equal to the total number of its connections with other nodes, $k_i = \sum_j  A(i,j)$. The In-degree of a node is equal to the number of edges connecting it to other nodes of the same community and can be calculated as $k_i^{in}(C) = \sum_{j\in C } A(i, j)$. The out-degree of a node $i$ is equal to the number of  connections to the nodes lying outside the community and can be calculated as $k_i^{out}(C) = \sum_{j \notin C} A(i,j)$.

\textbf{1. \textit{Mod Centrality:}} Masuda \cite{masuda} proposed to apply the dynamical importance strategy to the network representing the community structure. Given a community structure of the original network the community weighted network is build. Its nodes are the communities and the weight of a link between two communities is equal to the number of links they share. Then node that maximizes the following quantity are sequentially removed:
\begin{equation}
(2 \tilde{u}_K - x) \sum_{I \neq K}d_{kI}\tilde{u_I}
\end{equation}
The first factor measures the importance of the module that node $k$ belongs to (i.e. $(2 \tilde{u}_K - x) \approx 2 \tilde{u}_K$) where, $\tilde{u}_K$ represents the eigen vector corresponding to $K_{th}$ community. The second quantity in the equation represents the connectivity of node $k$ to other important modules (i.e., $\sum_{I \neq K}d_{kI}\tilde{u_I}$). 
On the basis of the modular structure determined for the original network, $\tilde{u}_I$ is recalculated and nodes are removed one at a time. If all the modules are isolated, nodes are removed sequentially in the descending order of in-degree, $k_i^{in}$ of the nodes. In-degree of all the remaining nodes is recalculated after the removal of each node. This strategy preferentially immunizes globally important nodes having important inter-community links rather than locally important ones such as local hubs (Algorithm \ref{a3}).

This strategy has got several drawbacks. It defines the importance of a node based on eigenvectors of the network. Calculation of eigenvectors is a computationally expensive task. Additionally, the centralities of the remaining nodes are re-calculated after the removal of nodes with high centrality. It also needs a coarse global information to which other community nodes are connected. When the communities are isolated after removing all the bridge nodes, it reduces to selecting the nodes only on the basis of higher in-degree. To overcome these drawbacks, we propose a strategy requiring less information, computationally cheaper and as efficient.

\begin{algorithm}
            \caption{To calculate Mod Centrality} \label{a3}

    \textbf{Input} {Graph/Network $G (V, E)$, Community Structure of the network}\\
    \textbf{Output} {A map/dictionary with $(node, centrality value)$ pairs}
    
    \begin{algorithmic}    
    
    \STATE Initialize an empty map, $M$ with (\textit{node, value}) pair format.
    \STATE Build a community weighted network using the community structure as follows: 
    \STATE Communities of the original network represent the nodes in the community weighted network.
    \STATE Total number of links between two communities denotes the weight of the link between the corresponding nodes in the weighted  network.
    \STATE Build an eigen matrix of the weighted network.
     \FOR {each node $k$ in the graph}
    
    \STATE Let $MC(k)$ be the mod centrality of node $k$.
    \STATE Let $\tilde{u}_K$ represents the eigen vector corresponding to the community of node $k$. $d_{kI}$ represents the number of inter-community links that exist between node $k$ and community $I$. 

     \IF {there are links remaining between any two communities in the network}
    \STATE 
    \begin{equation} 
    MC(k) = (2 \tilde{u}_K) \sum_{I \neq K}d_{kI}\tilde{u_I} \nonumber
     \end{equation}
        \ELSE
    \STATE $MC(k) = d_{kK}$ i.e. no of intra-connections of node $k$ in its own community
    
    \ENDIF
    \STATE Add the pair ($k$, $MC(k)$) to the map $M$.
    
    \ENDFOR
    \STATE Return the map $M$.
\end{algorithmic}
\end{algorithm}

\textbf{2. \textit{Proposed $Commn$ Centrality: }} $Commn$ centrality of a node is the weighted combination of its in-degree and out-degree. This centrality measure takes into account both intra and inter-community links of a node. It marks hubs or bridges as the influential nodes based on the strength of the community structure of the network. Let us consider an example of a network with community structure (Algorithm \ref{a4}).

\begin{algorithm}

    \caption{To calculate \textit{Commn} Centrality} \label{a4}
   \textbf{Input} {Graph/Network $G (V, E)$, Community Structure of the network with community list, $CL$ }\\
   \textbf{Output }{A map/dictionary with (node, centrality value) pairs}
    
    \begin{algorithmic}    
    
    \STATE Initialize an empty map, $M$ with (node, value) pair format.\\
    \FOR {each community $C$ in the list $CL$}
    
    \STATE calculate the fraction, $\mu_C$ of outer connections to the total connections in the community.
   
    \ENDFOR
   \FOR {each node $i$ in the graph}
 
    \STATE Calculate the no of intra-community links of node $i$, $k_i^{cin}$
    \STATE calculate the no of inter-community links of $i$, $k_i^{cout}$
    
        \ENDFOR
    \FOR {each community $C$ in the list $CL$}
    \FOR {each node $i$ in community $C$}
   \STATE  Calculate the Comm Centrality for node $i$ using the equation (reference to comm equation)
  
    \ENDFOR
    \STATE Add the pair ($i$, $CC(i)$) to the map $M$.
 
    \ENDFOR
    \STATE Return the map $M$.
    \end{algorithmic}
\end{algorithm}

\begin{figure}[htb!]
\begin{center}
\includegraphics[width=0.4\linewidth, height=2.2 in]{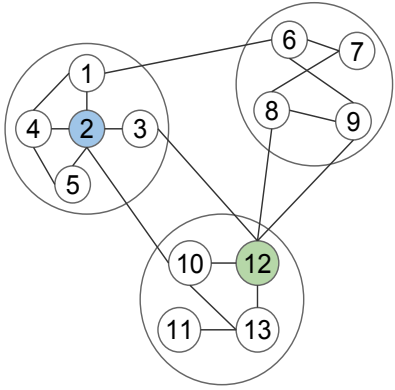} 
\end{center}
\caption{Proposed Centrality Measure} \label{f1}
\vspace{-1em}
\end{figure}

In Fig. \ref{f1}, it can be seen that node 2 has got many connections with the other nodes in its own community. It can be called as the hub or leader in its own community and thus can be labeled as an influential node. This is the node which will play a major role in spreading any kind of information or epidemics in its community. If this node is removed, there is very little chance of epidemic spreading to other nodes in the community. On the other hand, node 12 has got many connections with the nodes in other communities. It can be called as a bridge node, which connects its community to other communities in the network. Hence, it will also be responsible for information or epidemic propagation to other communities. So, it can be concluded that both community hubs and bridges are influential nodes in a network and play a key role in epidemic propagation. Hence, a centrality measure is needed which marks those nodes as influential which have got a good balance of inner and outer connections in the community. In other words, it must be successful in finding both the community hubs and bridges. We propose new centrality measure, $Commn$ centrality, defined as follows.

C$ommn$ centrality for a node $i$, belonging to community $C$ can be calculated as follows:
\begin{eqnarray}
CC(i)& = &(1 + \mu_C) * \left(\frac{k_i^{in}}{\underset{j}{max}\ (k_j^{in}\ \forall j \in C)} \ * \ R \right) + \nonumber \\
&& (1 - \mu_C) * \left(\frac{k_i^{out}}{\underset{j}{max}\ (k_j^{out}\ \forall j \in C)} \  * \ R \right)^2 \label{e3}
\end{eqnarray}

Here, $R$ is any integer number of choice, to get both in-degree and out-degree values in the same range i.e. $[0,\  R]$. As usually, out-degree values of the nodes are very low as compared to their in-degree values. A good choice for the value of $R$ will be maximum in-degree in the community $C$ i.e. $\underset{j}{max}\ (k_j^{in}\ \forall j \in C)$.

$\mu_C$ is the fraction of outer connections to the total connections in the community $C$, and can be calculated as follows:
\begin{equation}
\mu_C = \frac{\underset{i \in C}\sum k_i^{out}/k_i}{size(C)}
\end{equation}
\\
In equation \eqref{e3}, first in-degree and out-degree of the nodes are normalized by the maximum in-degree and out-degree, respectively in the community and brought in the same range. Then the term with out-degree is raised up to power two where as the term with in-degree is not raised. This is done to give more importance to out-degree of the nodes than in-degree while calculating their centrality value. Consider two nodes $i$ and $j$ belonging to the same community. If we have to mark one node as more influential than the other, then there can be three possible cases to consider. 
\begin{itemize}
  \item First case, both in-degree and out-degree of one node, say $i$ is larger than the other node $j$ i.e.
  \begin{equation}\nonumber
 k_i^{in} > k_j^{in} \ \ and  \ \ k_i^{out} > k_j^{out}
   \end{equation}
  \item Second, in-degree (or out-degree) of node $i$ is equal to that of node $j$, and out-degree (or indegree) of $i$ is larger than that of $j$ i.e.
   \begin{equation}\nonumber
	( k_i^{in} = k_j^{in} \ and \  k_i^{out} > k_j^{out} ) \ \ or \ \ ( k_i^{out} = k_j^{out} \  and \  k_i^{in} > k_j^{i1n} )
   \end{equation}
  \item Third, indegree of one node, say $i$ is larger than that of other node $j$, but $j$ has got higher outdegree than node $i$.
   \begin{equation}\nonumber
 	k_i^{in} > k_j^{in} \ \ and  \ \ k_i^{out} < k_j^{out}
   \end{equation}
\end{itemize}

In the first two cases, indisputably, node $i$ will be marked as more influential. But, in the third case, we mark node $j$ as more central or influential than node $i$ because of more outer connections of $j$. The only constraint is that the in-degree of node $j$ must not be very smaller than that of $i$. That is, node $j$ must also have comparable number of inner connections. Specifically, as per equation \eqref{e3}, node $j$ is marked more influential if $(k_i^{in} - k_j^{in}) \leq ((k_j^{out})^2 - (k_i^{out})^2)$, otherwise node $i$ is considered more influential.

Also note, in the equation \eqref{e3}, $\mu_C$ and $(1 - \mu_C)$ are assigned as weights to the in-degree and out-degree terms respectively. This is done to adapt to the strength of a particular community. For example, if a community is very cohesive, i.e. there are very less outer connections as compared to total connections. $\mu_C$ value of this community will be low and hence more weight will be given to the out-degree terms to target the rare nodes with outer connections in the community. On the other hand, if a community has a lot of outer connections and accordingly large value of $\mu_C$, more weight will be given to the in-degree term.
 
Nodes are removed sequentially from each community in the decreasing order of their centrality value in their respective community. Number of nodes to be removed from a community are kept proportional to the community size. Hence, more nodes are removed from larger communities than from the smaller communities. Also note that removing a node can modify the network topology and hence the centrality of the remaining nodes might not be the same. So, the centralities of the remaining nodes are recalculated after removing the node with highest centrality value.

\section{Classical epidemic spreading, SIR model} \label{s3}

\hspace*{5mm}
Classical SIR model \cite{morsat,Newep} is one of the most investigated epidemic spreading model for complex networks. In this model, nodes are in one of the three compartments \textemdash \textbf{S}usceptible (healthy), \textbf{I}nfected (those who infected with disease and also actively spread it) and \textbf{R}emoved (immunized or dead). Susceptible nodes represent the individuals which are not yet infected with the disease. Infected nodes are the ones which are infected with the disease and can spread the disease to the susceptible nodes. Removed nodes represent the individuals which are infected and are immunized or died. These nodes are not neither infected again, nor they can transmit the infection to others. The epidemic is propagated through the nodes by pairwise contacts between the infected and susceptible nodes in the network. Following the law of mass action, the spreading process evolves with direct contact of the infected nodes with susceptible nodes in the population. The epidemic spreads from one infected node to another susceptible nodes in a single time step with $\lambda$ rate, if an undirected edge exists between them. An infected node become removed node if it is immunized/dead with rate $\sigma$. 
Here, $\mathbb{S}(k,t), \mathbb{I}(k,t), \mathbb{R}(k,t)$ are the expected values of susceptible, infected and removed nodes in network with degree $k$ at time $t$. Let $S(\emph{k,t})=\mathbb{S}(k,t)/N(k), I(\emph{k,t})=\mathbb{I}(k,t)/N(k), R(\emph{k,t})=\mathbb{R}(k,t)/N(k)$
be the fraction of susceptible, infected and removed
nodes, respectively with degree \textit{k} at time \textit{t}. These fractions of
the nodes satisfy the normalization condition, $
S(\emph{k,t})+I(\emph{k,t})+R(\emph{k,t})=1$, where $N(k)$ represents the total number
of nodes with degree $k$, in the network.
Above epidemic spreading process can be summarized by the following set of pairwise
interactions.

\begin{eqnarray*}
&&S_1+I_2\xrightarrow{\lambda}I_1 + I_2,\\
&&\mbox{(when infected meets with the susceptible, it makes them infected} \\ &&\mbox{at rate $\lambda$)}\\
&&I_1\xrightarrow{\sigma}R_1,\\
&&\mbox{(spontaneously any infected node will become removed node with rate $ \sigma $}
\end{eqnarray*}

The formulation of this model for analyzing complex networks as interacting Markov chains. The framework is used to derive from the first-principles, the mean-field equations for the dynamics of epidemic spreading in the uncorrelated heterogeneous complex networks (scale free networks) with arbitrary
correlations. These are given below.
 \begin{eqnarray}
\frac{dS(\emph{k,t})}{dt}&=&-k\lambda S(\emph{k,t})\sum_{l}I(l,t)P(l|k).\label{1-1}\\
\frac{dI(\emph{k,t})}{dt}&=& -\sigma I(\emph{k,t})+\lambda S(k,t) \sum_{l} I(l,t)P(l|k).\label{1-2}\\
\frac{dR(\emph{k,t})}{dt}&=& \sigma I(\emph{k,t}).\label{1-3}
\end{eqnarray}

Where the conditional probability $P(l|k)$ is the degree-degree correlation function that is randomly chosen edge emanating from a node of degree \emph{k} leads to a node of degree \emph{l}. Here, it is assumed that the degree of nodes in the whole network are uncorrelated. Therefore, degree-degree correlation is $ P(l|k)=\frac{lP(l)}{ \langle k\rangle}$ for uncorrelated complex networks where $P(l)$ is the degree distribution and  $\langle k\rangle$ is the average degree of the network (the edge will be biased to fall on vertices of high degree, hence the conditional probability $P(l|k)$ is proportional to $kP(k)$). Epidemic spreading model is defined for correlated scale free networks \cite{anuncc}. It is shown that the critical threshold for epidemic spreading is independent of the removing mechanism. It was found that the critical threshold as, $\lambda_c=\frac{\langle k\rangle}{\langle k^2\rangle}$. Hence, it implies that epidemic threshold is absent in large size scale free networks $(\langle k^2\rangle\rightarrow \infty, \lambda_c\rightarrow 0)$. This result is not good for epidemic control,  since the epidemics will exist in the real networks for any non zero value of spreading rate $\lambda$.

\section{Synthetic Benchmark Data} \label{s4}
Generative models allow to produce large variations of synthetic networks easily and quickly. These models provide a control on some topological properties of the generated networks, making it possible to mimic the targeted system features. The only point of concern is the level of realism of the generated networks, which is a prerequisite to obtain relevant test results. To date, the LFR (Lancichinetti, Fortunato and Radicchi) \cite{LFR} model is the most efficient solution in order to generate synthetic networks with community structure. Consequently, it is used to generate the networks with a non-overlapping community structure. It is based on the configuration model (CM) \cite{CM}, which generates networks with power law degree distribution. The generative process performs in three steps. First, it uses the configuration model to generate a network with a power-law degree distribution with exponent $\gamma$. Second, virtual communities are defined so that their sizes follow a power-law distribution with exponent $\beta$. Each node is randomly affected to a community, provided the community size is greater or equal to the node internal degree. Third, an iterative process takes place in order to rewire certain links, such that the proportion of intra-community and inter-community links is changed so that it gets close to the mixing coefficient value $\mu$, while preserving the degree distribution. This model guarantees to obtain realistic features (power-law distributed degrees and community sizes). It also includes a rich set of parameters which can be tuned to get the desired network topology. These parameters are- the mixing parameter $\mu$, the average degree $k$, the maximum degree $k_{max}$, the maximum community size $c_{max}$, and the minimum community size $c_{min}$.

For small $\mu$ values, the communities are distinctly separated because they share only a few links, whereas when $\mu$ increases, the proportion of inter-community links becomes higher, making community identification a difficult task. The network has no community structure for a limit value of the mixing coefficient given by: 
\begin{equation}
\mu_{\lim} > (n - n_c^{max})/n,
\end{equation} 
where $n$ and $n_c^{max}$ are number of nodes in the network and in the biggest community, respectively \cite{LFb}.

\section{Experimental Set-up} \label{s5}
To investigate the spread of an infectious disease, the classical SIR model of epidemic spreading is used. Initially, all the nodes are treated as susceptible. After this initial set-up, a fraction of  nodes are chosen at random to be infected and the remaining nodes are considered to be susceptible (Algorithm \ref{a5}).

\begin{algorithm}
  \caption{SIR Algorithm} \label{a5}

    
    \textbf{Input}:{Graph $G(V, E), I$, $\lambda$, $\sigma$} \\
    \textbf{Output}:{Total no of nodes infected during the process, $T\_I$}
    
    \begin{algorithmic}    
    
    \STATE  Select $I$ number of nodes at random to be marked as infected and place them in infected\_list.
    \STATE  Mark all other nodes as susceptible.
    
    \WHILE{infected\_list is not empty}
        \STATE  select one infected node $u$ from the list.
  \FOR {every node $v$ adjacent to $u$}
  
    \IF{$v$ is susceptible}
      
        \STATE with probability equal to $\lambda$: 
        \STATE mark $v$ as infected, and add to the infected\_list.
        \STATE increase $T\_I$ by one.
      
     \ELSE
      
        \STATE with probability equal to $\sigma$:
        \STATE mark $v$ as recovered, and remove from the infected\_list.
      
    \ENDIF
      
    \ENDFOR
  
     \ENDWHILE
    \RETURN $T\_I$

  \end{algorithmic}
\end{algorithm}

The infection spreads through the contact network during each time step, where $\lambda$ is the transmission rate of infection being spread from an infected node to a susceptible node, and $I$ is the number of infected neighboring nodes. Infected nodes removed with rate $\sigma$ at each time step. 
After the recovery of a node occurs, its state is toggled from infected to resistant. Simulations are halted if there are no infected nodes left in the network, and the total number of infected nodes is analyzed. This time foe halting is called steady state time.

To investigate the efficiency of the proposed strategies, synthetic networks are generated using the LFR algorithm. In order to select appropriate values for the network parameters, we considered several studies of real-world networks. For the power-law exponents, we used $\gamma = 3$ and $\beta = 2$, which seem to be the most representative values. Concerning the number of nodes and links, no typical values emerge. Studies show that real-world complex networks can have very different sizes, defined on a wide range going from tens to millions of nodes. The average and maximal degrees are also very variable, so it is difficult to characterize them too. As a result, we selected some consensual values for these parameters, while considering also the computational aspect of the simulations \cite{Hocine}. The parameters values used for the generation of the LFR networks in this study are given in the Table \ref{t2}.

\begin{table}[htb!]
\caption{Parameters for the LFR network generation} \label{t2}
\centering
\begin{tabular}{| c | c |} \hline
\hline
Number of nodes, n & 7500 \\ \hline
Average degree, $\langle k \rangle$ & 10 \\ \hline
Maximum degree, $k_{max}$ & 180 \\ \hline
Mixing parameter, $\mu$ & 0.2, 0.3, 0.5 \\ \hline
$\gamma$ & 3 \\ \hline
$\beta$ & 2 \\ \hline
Minimum community size, $C_{min}$ & 5 \\ \hline
Maximum community size, $C_{max}$ & 180 \\ \hline
\end{tabular}
\end{table}

\begin{figure}[htb!]
\begin{center}
$\begin{array}{cc}
\includegraphics[width=0.5\linewidth, height=2.4 in]{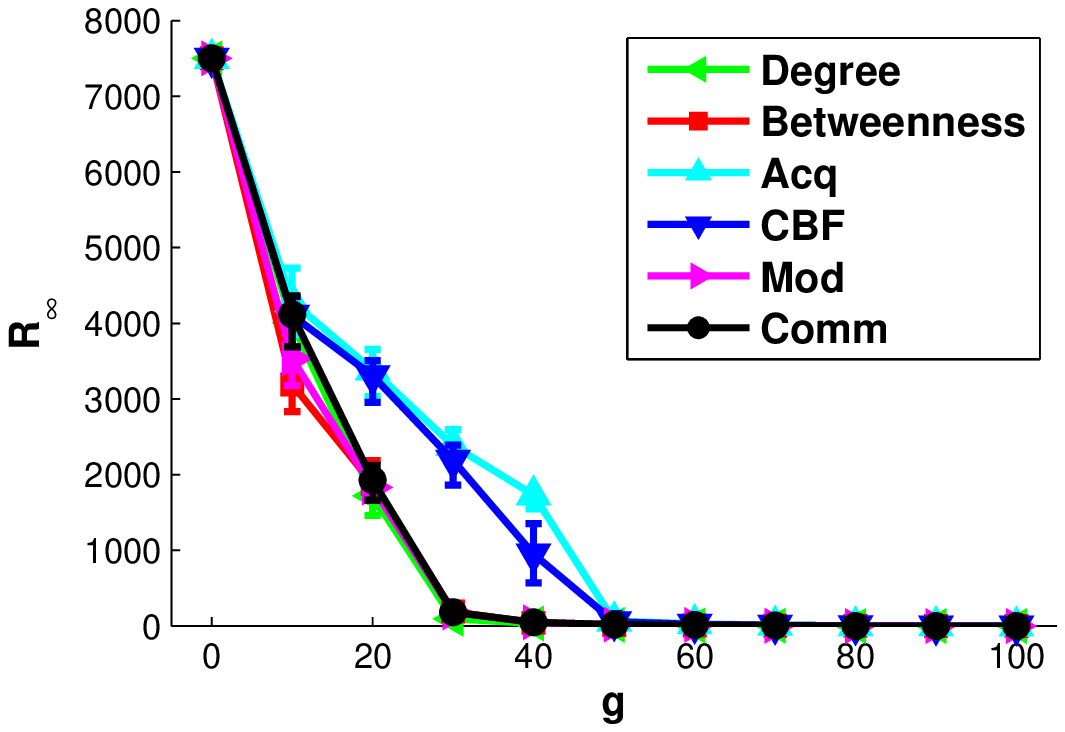} &
\includegraphics[width=0.5\linewidth, height=2.4 in]{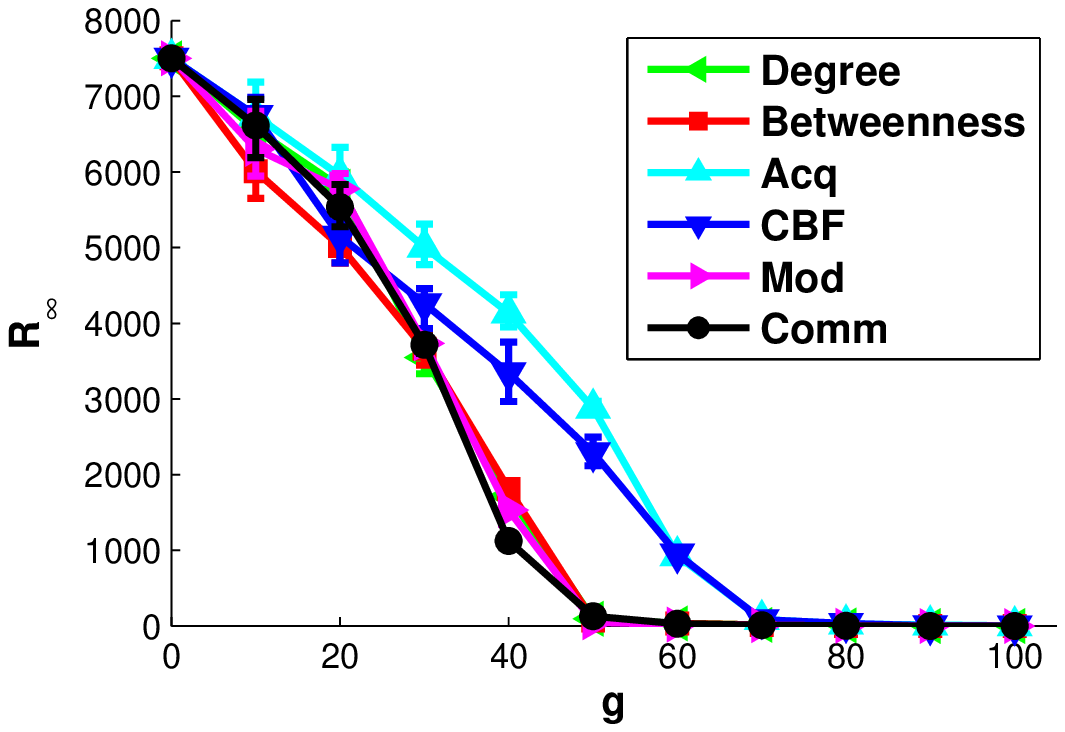} \\
\mbox{(a) $\mu$=0.2, $\lambda$=0.1} & \mbox{(b) $\mu$=0.2, $\lambda$=0.9} \\
\includegraphics[width=0.5\linewidth, height=2.4 in]{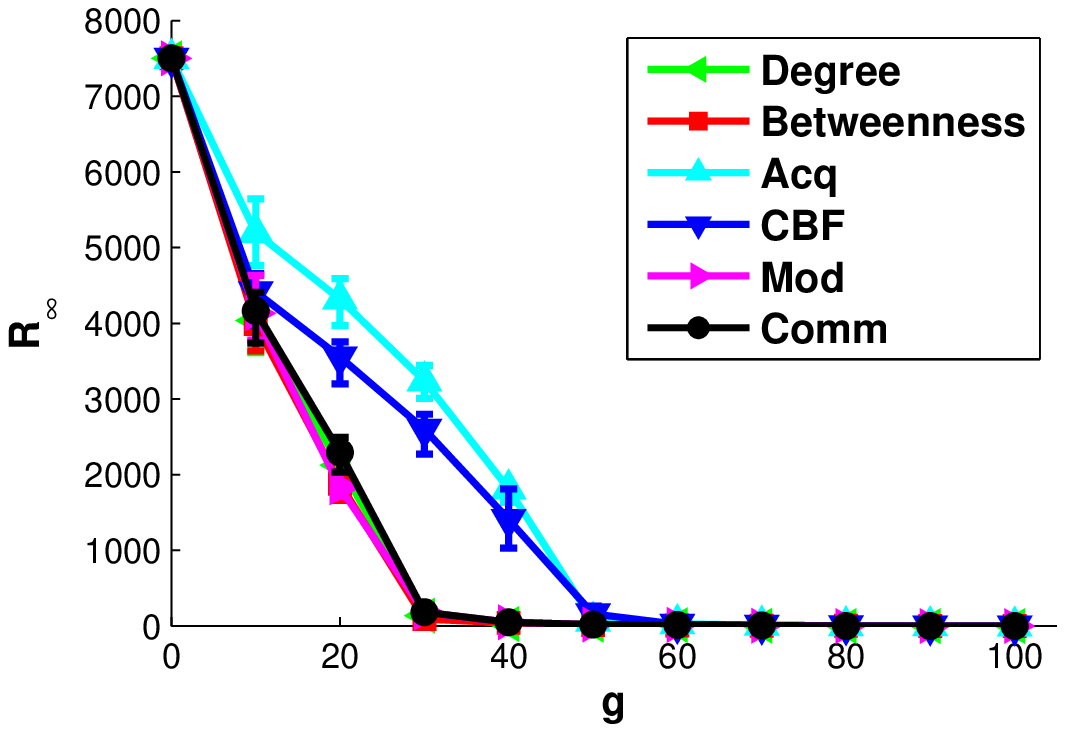} &
\includegraphics[width=0.5\linewidth, height=2.4 in]{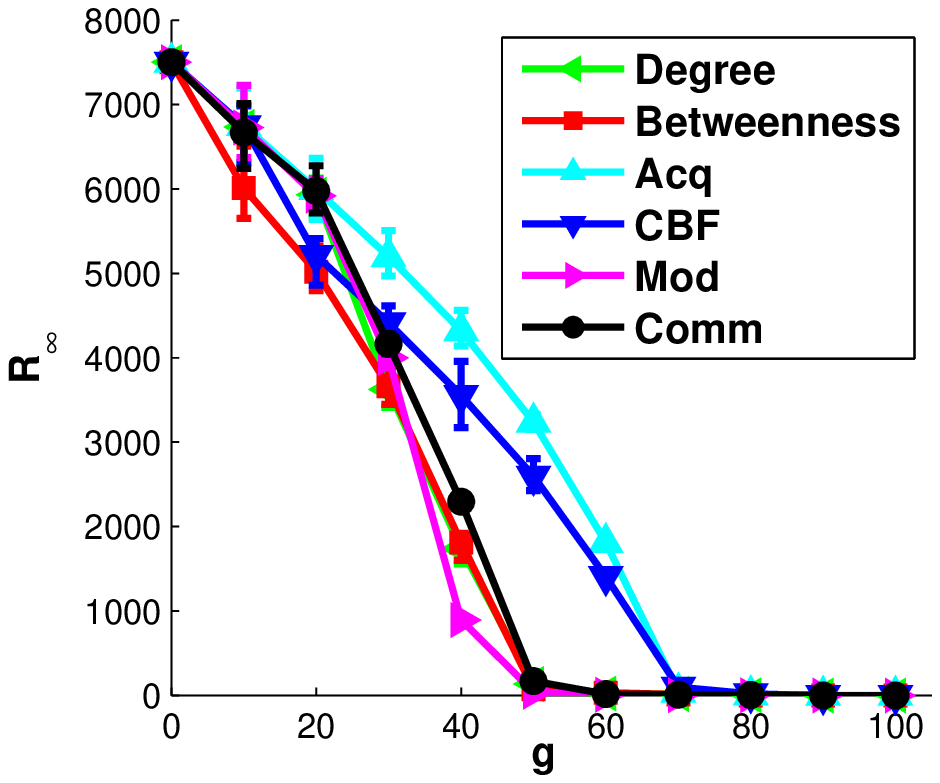} \\
\mbox{(c) $\mu$=0.3, $\lambda$=0.1} & \mbox{(d) $\mu$=0.3, $\lambda$=0.9} \\
\includegraphics[width=0.5\linewidth, height=2.4 in]{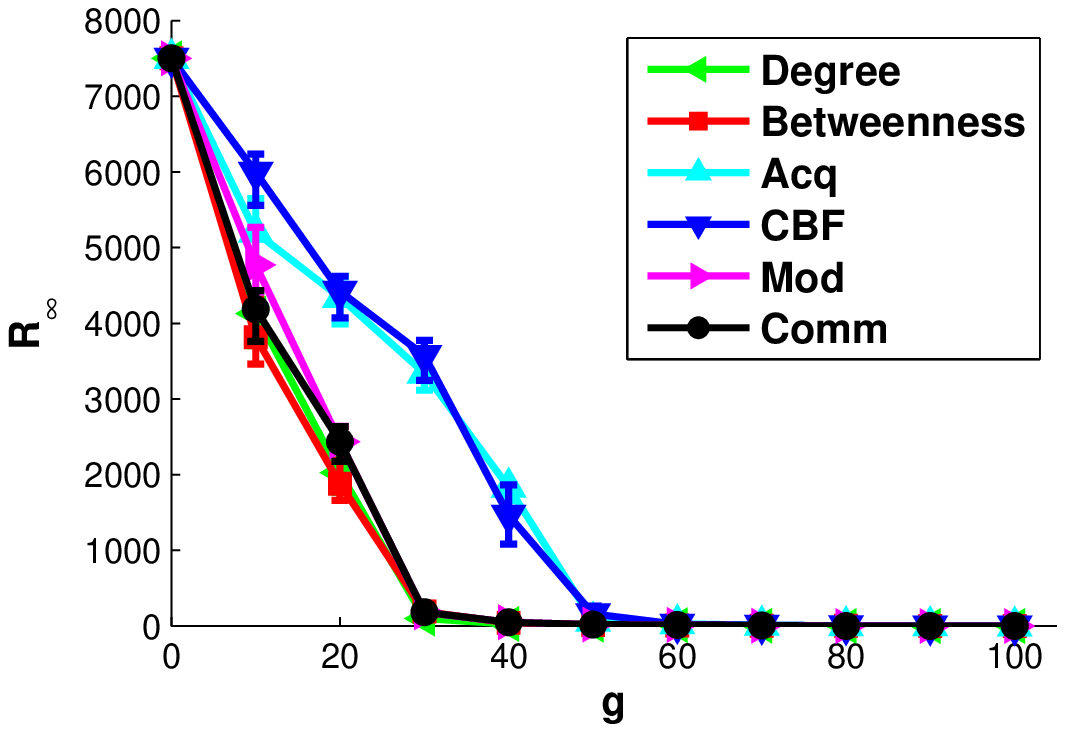} &
\includegraphics[width=0.5\linewidth, height=2.4 in]{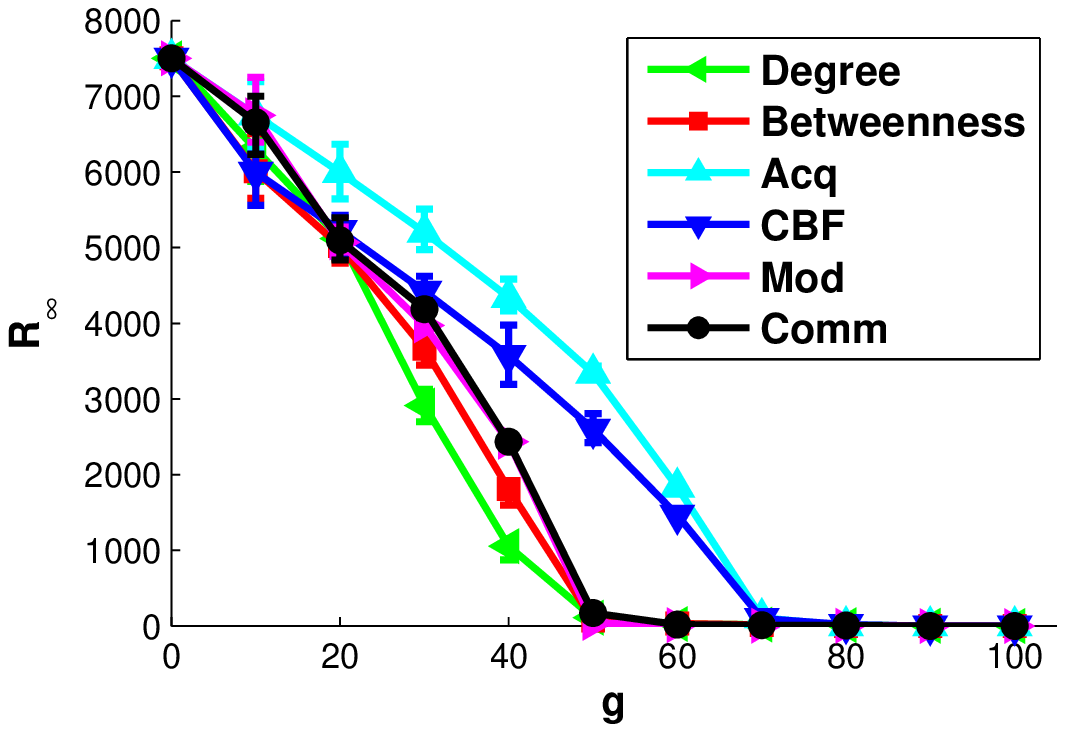}\\
\mbox{(e) $\mu$=0.5, $\lambda$=0.1} & \mbox{(f) $\mu$=0.5, $\lambda$=0.9} 
\end{array}$
\end{center}
\caption{Effect of various immunization strategies on the total number of infected nodes during the SIR simulation on LFR network with $\sigma$ = 0.1} \label{f2}
\vspace{-1em}
\end{figure}

In order to better understand the influence of the community structure, networks with various $\mu$ values (ranging from 0.2 to 0.5) are generated. For each $\mu$ value, 10 sample networks are generated. Each simulation is run on these 10 networks and the mean values of the results together with their standard deviation are reported in the Fig. \ref{f2}.

After forming the community structure various immunization strategies are performed on the given network according to the Algorithm \ref{a6}. The network obtained after applying immunization is used to study the epidemic diffusion by using SIR model.

\begin{algorithm}
   \caption{Immunization Algorithm} \label{a6}
       
    \textbf{Input}:{Graph $G(V, E)$, Centrality measure($C$), No of nodes to 
immunize($n$)}
    \textbf{Output}:{Graph with Immunized/Removed Nodes}
    
    \begin{algorithmic}    
    
    \STATE 1. Calculate the centrality of every node in the graph using the measure $C$.
    \STATE 2. Sort the nodes in decreasing order of their centrality values.
    \STATE 3. Remove top $n$ nodes with highest centrality values from the graph $G$.
    \STATE 4. Return graph $G$ with immunized/removed nodes.
    
\end{algorithmic}
\end{algorithm}

\begin{figure}[htb!]
\begin{center}
$\begin{array}{cc}
\includegraphics[width=0.5\linewidth, height=2.4 in]{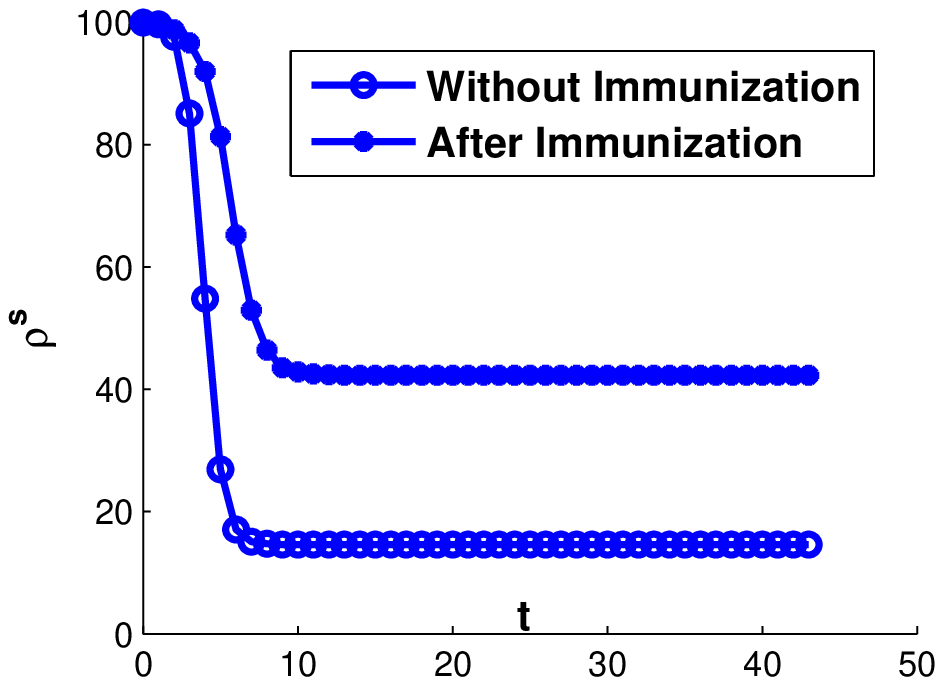} &
\includegraphics[width=0.5\linewidth, height=2.4 in]{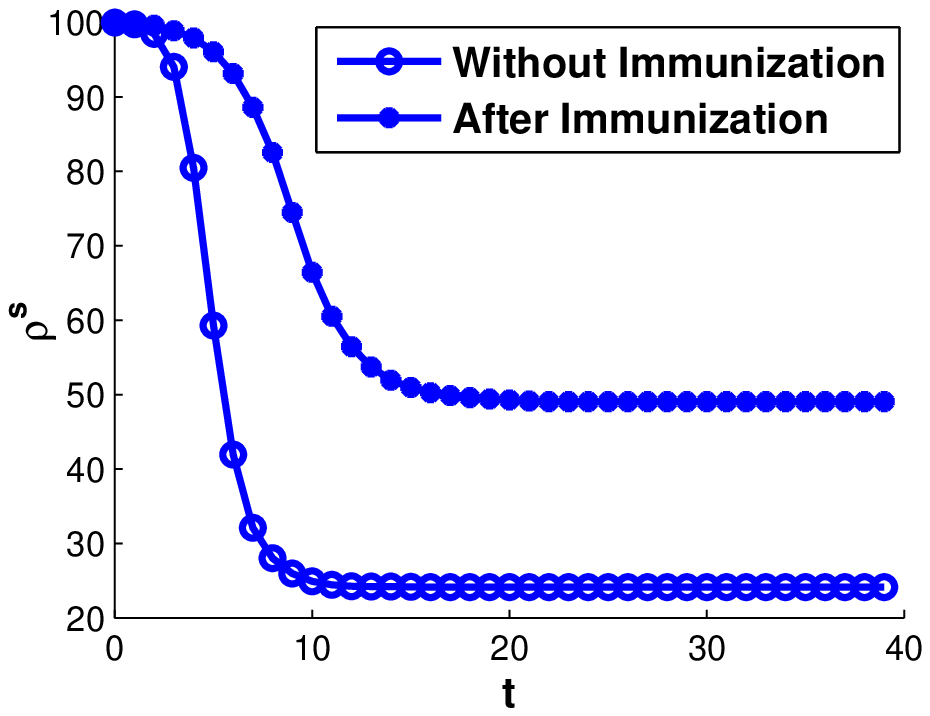} \\
\mbox{(a) $\lambda=1$ and $ \sigma=0.1 $} & \mbox{(b) $\lambda=0.5$ and $ \sigma=0.1 $} \\
\includegraphics[width=0.5\linewidth, height=2.4 in]{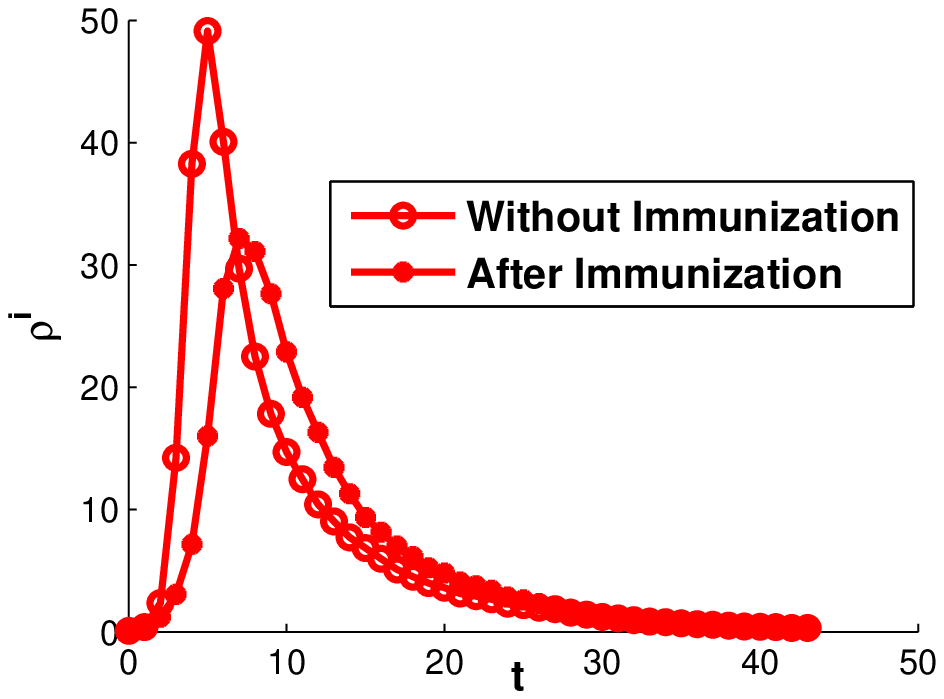} &
\includegraphics[width=0.5\linewidth, height=2.4 in]{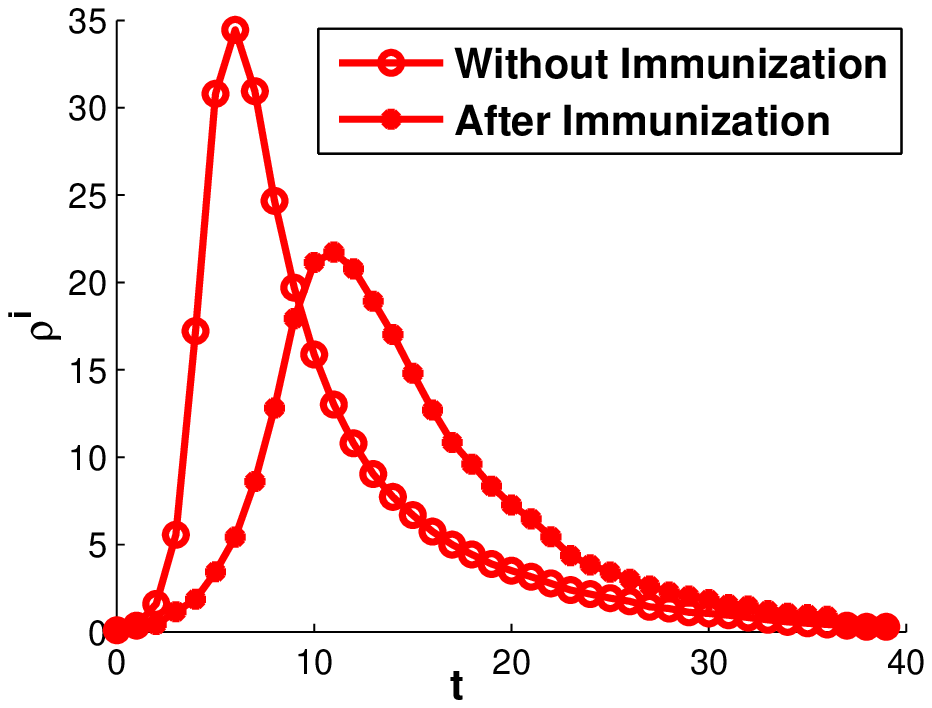} \\
\mbox{(c) $\lambda=1$ and $ \sigma=0.1 $} & \mbox{(d) $\lambda=0.5$ and $ \sigma=0.1 $} \\
\includegraphics[width=0.5\linewidth, height=2.4 in]{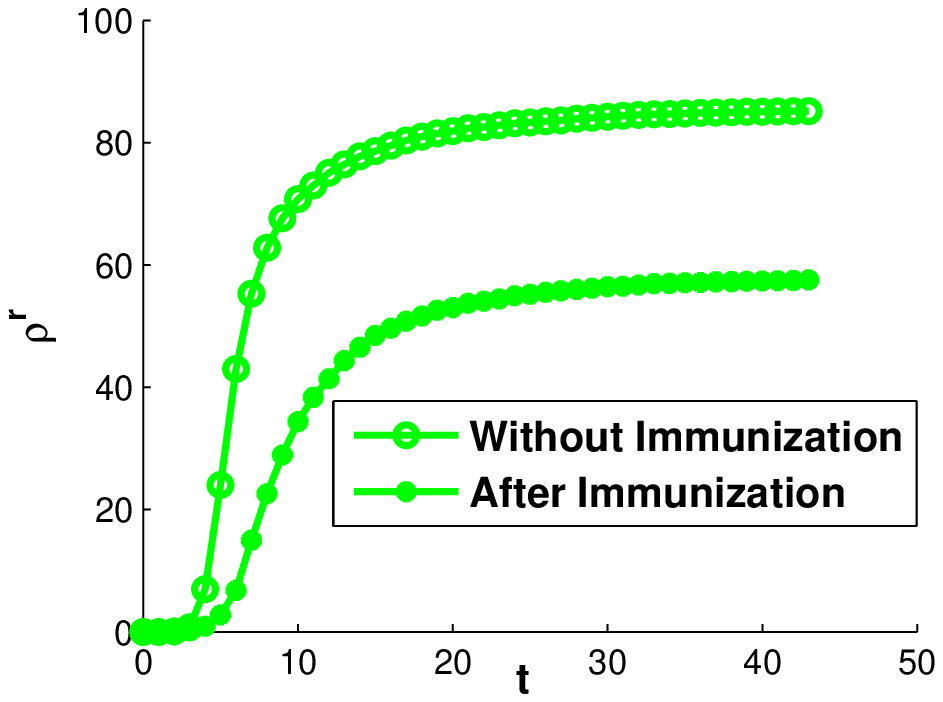} &
\includegraphics[width=0.5\linewidth, height=2.4 in]{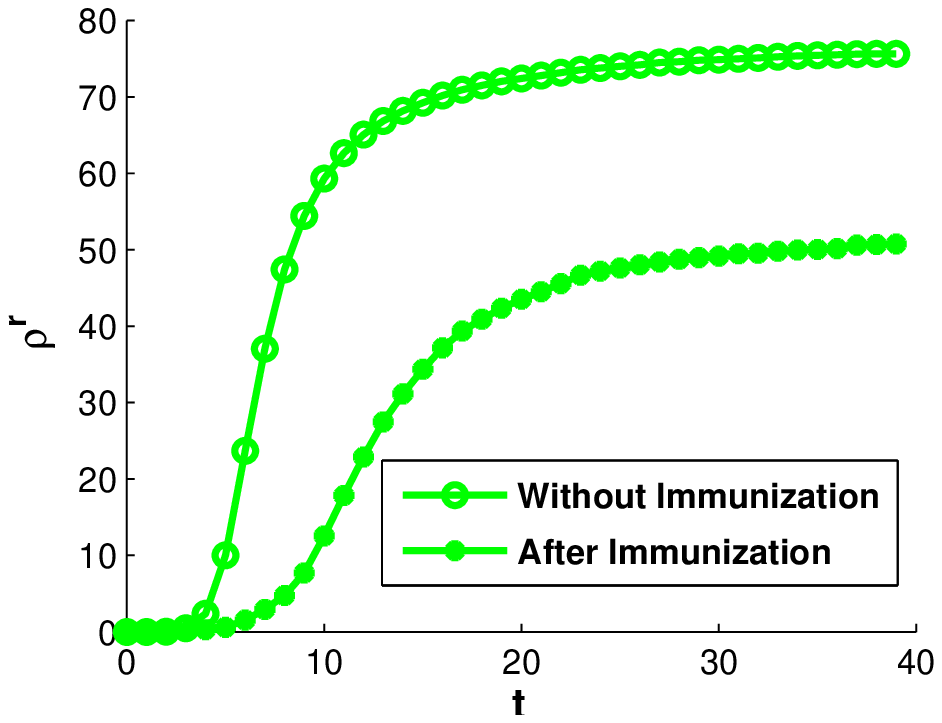}\\
\mbox{(e) $\lambda=1$ and $ \sigma=0.1 $} & \mbox{(f)$\lambda=0.5$ and $ \sigma=0.1 $} 
\end{array}$
\end{center}
\caption{Dynamical diffusion over the network against time, $ t $. Fraction of the susceptible, infected and removed nodes (\%) in the population respectively, $\rho^s$, $\rho^i$ and $\rho^r$ nodes without immunization and with immunization (20\%) strategy for different values of $ \lambda $ and $ \sigma $} \label{f21}
\end{figure}

\section{Results and Discussion}
In the simulations, final population of the removed nodes ($ R_\infty $) is calculated to study the epidemic outbreak in the given population (network). As, $ R(k,t) $ represents the size of removed nodes with degree $ k $ at time $ t $. Hence, the final number of removed nodes at $ t=\infty $, time, $ \infty $ is sufficiently large at which epidemic process will be achieved its steady state. At this time no more infected nodes will left. The total number of remove nodes are
\begin{eqnarray}
&&R_t=\sum_k{R(k,t)} \nonumber \\
\mbox{at $ t=\infty $} \\
&&R_\infty=\sum_k{R(k,\infty)} \nonumber 
\end{eqnarray}

The results of the simulations using various available immunization strategies (degree, betweenness, acquaintance, CBF, mod) and proposed immunization strategies, \textit{Commn} on LFR generated networks are shown in Fig. \ref{f2}. Here, recovering rate $\sigma$ is kept equal to 0.1 and spreading rate, $\lambda$ is varied with two values, 0.1 and 0.9. The removed fraction of nodes (\%) is mentioned by $ g $. Let us take the case when $\lambda = 0.1$ at the left side of the Fig. \ref{f2} for all $\mu$ values. The global centrality based methods (degree, betweenness) work efficiently in every case. After removing or immunizing only 30\% of the nodes and epidemic spreading die. This is expected as both the methods exploit the information about the overall network topology. When the stochastic strategies (Random Acquaintance and \textit{CBF}) are used, 50\% nodes need to be removed in order to stop the epidemic spreading. This is the price to pay for not knowing the global structure of the network. It is interesting that the proposed community-based centrality measures, $Commn$ and $Mod$ strategies are almost as effective as the global centrality methods despite the fact that they are agnostic about the full network structure. Furthermore, they perform a lot better than the Acquaintance and \textit{CBF} strategies, with just the information about the community structure of the network.

The reason why the proposed $Commn$ centrality based strategy is effective, can be explained by the specific position of the targeted nodes within the network. It takes into account, both the in-degree and out-degree of the nodes. First it tries to remove or immunize the nodes which have more outer connections to other communities. In the equation \ref{e3}, out-degree of the node is raised up to power two, for this purpose. In a group of people, these are the individuals which pass all the information contained in their group to other groups. These nodes can be considered to be the bridges between the communities. When these nodes are removed, most of the paths or bridges between different communities are lost. Communities are isolated and thus the epidemic is not able to spread across the communities. After removing the nodes with higher out-degrees, nodes are selected which have lot of connections inside their own community. They can be considered as the core points of their community. In a network of people, these individuals are the leaders, representatives or agents of information flow in their communities. These high in-degree nodes are connected to many other nodes in their community. Most of the regular nodes in a community are not directly connected to each other. They would be connected to each other through the paths which would most likely contain these high in-degree nodes. These nodes if infected from outside, have a greater chance of infecting the whole community. When these high in-degree nodes are removed from the communities, the communities break from inside. In other words, the paths connecting the regular nodes to each other are broken. The remaining nodes are not able to contact each other and thus the epidemic is not able to affect a significant part of community, and it dies soon.

The underlying idea in proposing the centrality measure which is simply based on in-degree and out-degree of the nodes is that they intuitively represent the global degree and betweenness centralities at the community level. A node with a high in-degree or out-degree  has generally a high overall degree. Indeed, the total degree of a node is the addition of  in-degree and out-degree. A node with high in-degree is probably a node with a high betweenness measure in its community, as it will be contained in most of the paths connecting the regular nodes to each other in the community. Similarly, a node with high out-degree is probably a node with high betweenness measure in the overall network. Indeed, these high out-degree nodes are part of most of the paths connecting the nodes falling in different communities.

Fig. \ref{f2} \{(b), (d), (f)\} reports the SIR simulation results for $\lambda$ = 0.9 and $\sigma$ = 0.1. The results are very similar except the fact that in this case, a greater $ g $ need to be immunized in order to stop the epidemic spreading. This is true for all the strategies. For example, when $\lambda$ is equal to 0.1, degree and betweenness centrality based methods required only 30\% of the nodes to be removed to mitigate the epidemic spreading, whereas now they require 50\% of the nodes to be immunized or removed. Indeed, when $\lambda$ increases, the probability of an infected node to infect its neighbors gets higher. So, the epidemic spread at a higher rate, and thus more nodes are needed to be immunized to prevent the epidemic spreading.

The dynamics of the fraction of   susceptible ($ \rho^s $), infected ($\rho^i$) and removed ($\rho^r$ )  nodes (in \%) for different values of spreading rate $ \lambda $and removing rate $ \sigma $ are reported  in Fig. \ref{f21} plots without immunization and with the proposed community based immunization by considering, $g = 20\%$ are given for comparative purposes.

The dynamics of  of respectively, the fraction of nodes of the total nodes in \% for the susceptible ($ \rho^s $), infected ($\rho^i$) and removed ($\rho^r$ ) for different values of spreading rates, $ \lambda $ and removing rate $ \sigma $ is reported in Fig. \ref{f21}. In the given Fig. \ref{f21}, time evolution is plotted without immunization and with proposed community based immunization, \textit{Commn}, strategy.

\begin{figure}[htb!]
\begin{center}
$\begin{array}{cc}
\includegraphics[width=0.5\linewidth, height=2.4 in]{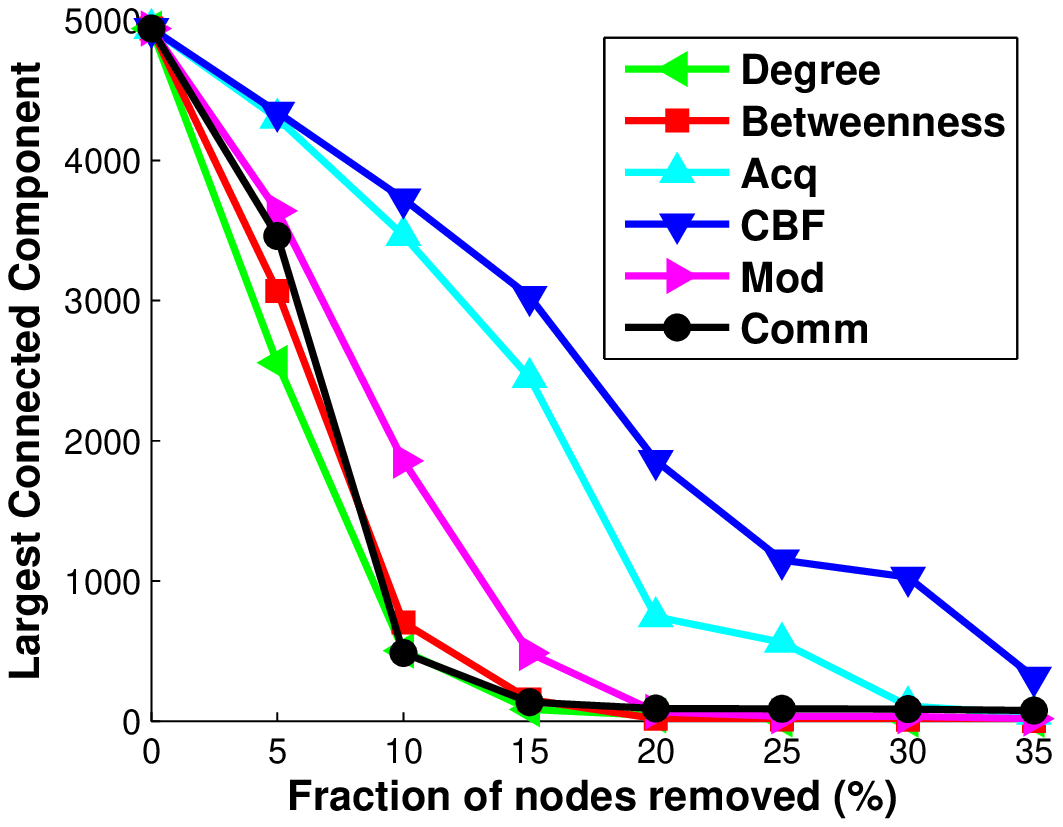}  &
\includegraphics[width=0.5\linewidth, height=2.4 in]{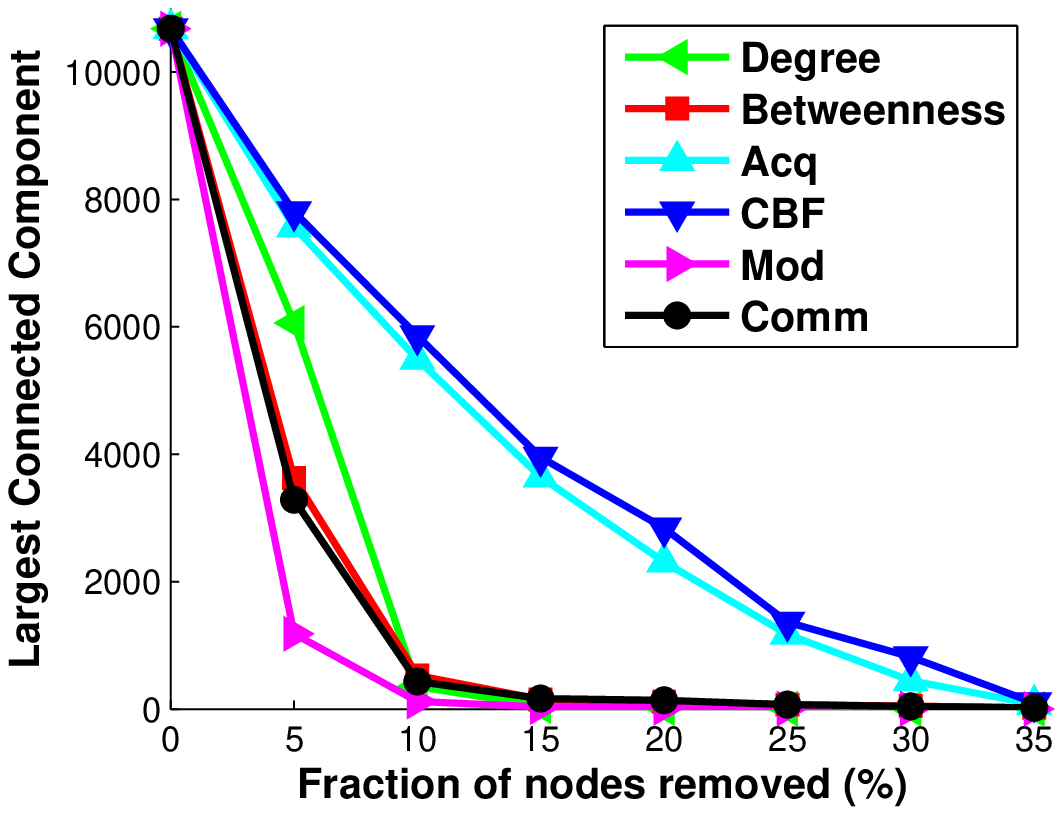}  \\
\mbox{(a) Power Grid Network} & \mbox{(b) PGP network} \\
\includegraphics[width=0.5\linewidth, height=2.4 in]{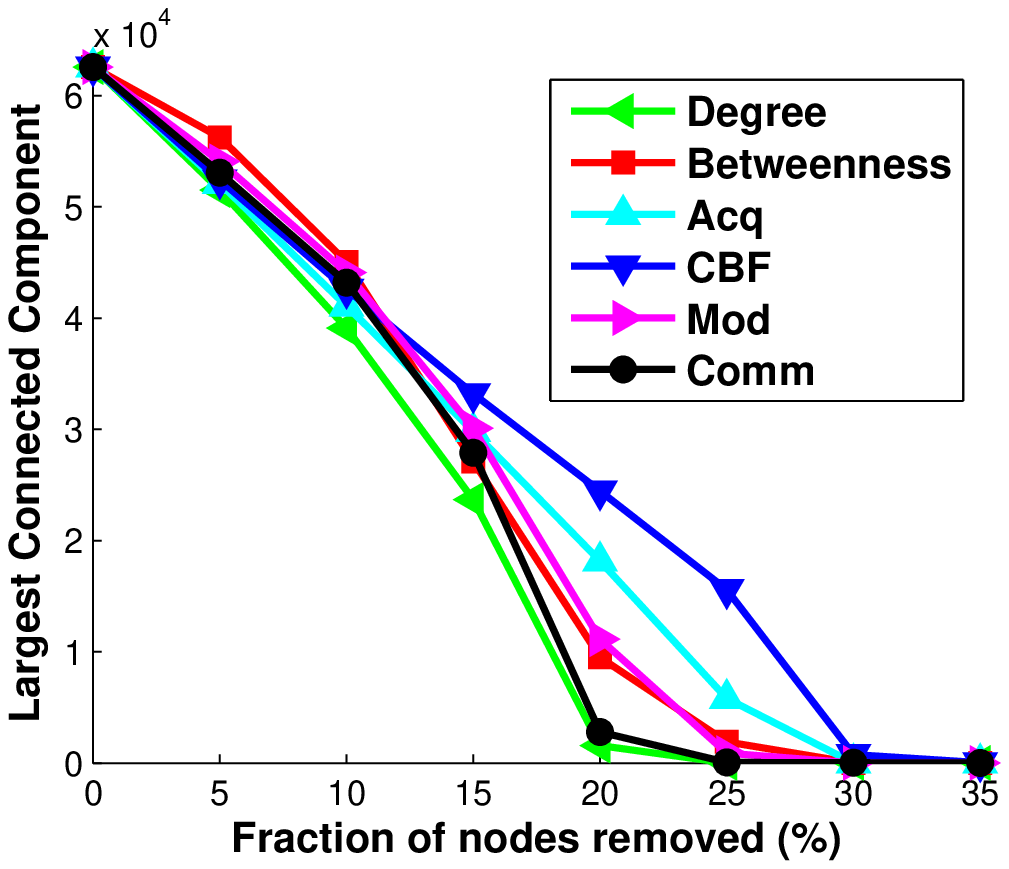}  &
\includegraphics[width=0.5\linewidth, height=2.4 in]{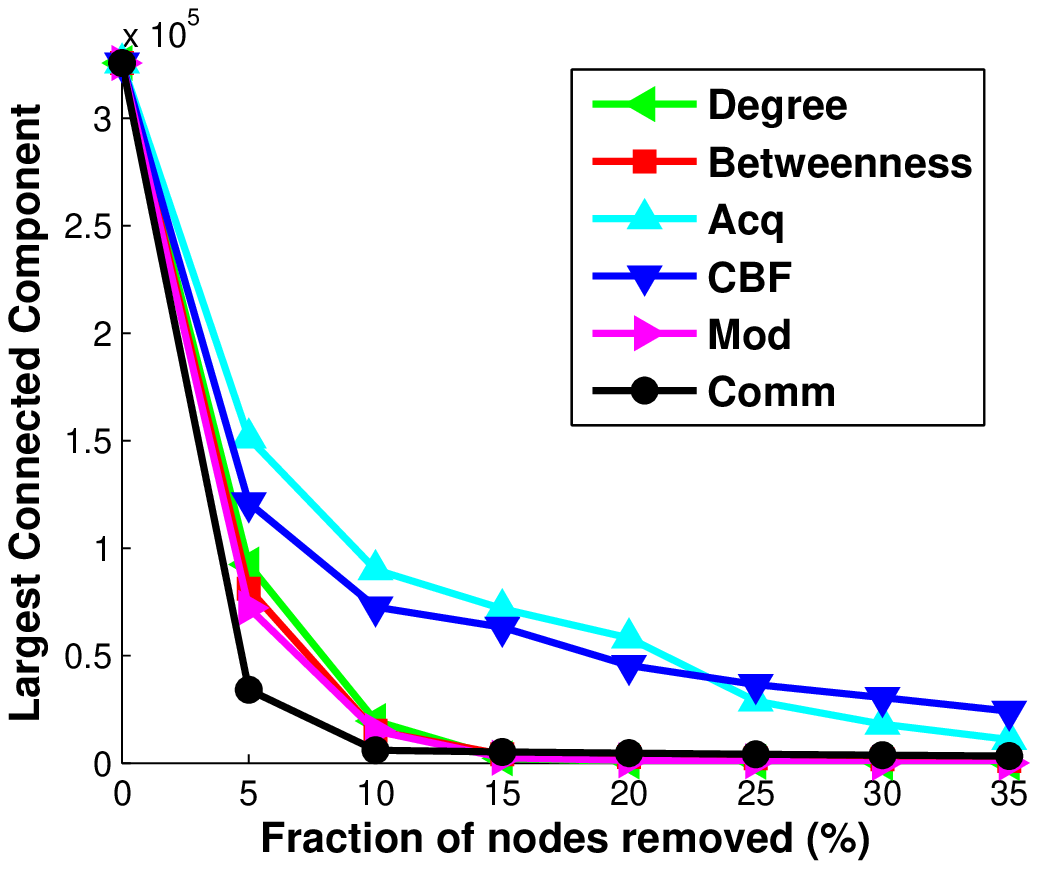} \\
\mbox{(c) Gnutella network} & \mbox{(d) WWW network} 
\end{array}$
\end{center}
\caption{Size of Largest Connected Component against the various centrality measures (a) Power Grid Network (b) PGP Network (c) Gnutella Network (d) WWW Network}\label{f3} 
\vspace{-1em}
\end{figure}

We also used real-world networks to further validate the results on synthetic networks with controlled properties. Statistics of these networks are given in Table \ref{t3}. Community structures of the networks are discovered using local community detection algorithms \cite{chen, martelot}. Figs. \ref{f3}(a)-\ref{f3}(d) show the results of comparison of available and proposed immunization strategies on these real-world networks.

\begin{table}[htb!]
\caption{Statistics of Real-World Networks} \label{t3}
\centering
\begin{tabular}{| c | c | c | c |}
\hline
\textbf{Dataset} & \textbf{Nodes} & \textbf{Edges} & \textbf{No. of Communities}\\ \hline
Power Grid Network & 5,004 & 6,597 & 168 \\ \hline
PGP Network & 10,680 & 24,316 & 153 \\ \hline
Gnutella Network & 62,586 & 1,47,892 & 171 \\ \hline
WWW Network & 3,52,729 & 11,17,563 & 3060 \\ \hline
\end{tabular}
\end{table}

The first network is the Power grid network \cite{sw}. Power grid data set comprises of an undirected, unweighted network representing the topology of the Western States Power Grid of the United States. Numerical results are shown in Fig. \ref{f3}(a). Proposed $Commn$ centrality performs as well as degree centrality and better than all other centrality measures.

The Second example is the social network called Pretty Good Privacy(PGP) network \cite{pgp}. This is the interaction network of users of the PGP algorithm. A link between two persons denotes the sharing of confidential information using the encryption algorithm on the internet. Results for the network are shown in Fig. \ref{f3}(b). Here, Mod centrality works a little better than the $Commn$ centrality, but at least both takes around 10\% of the nodes needed to be removed from the network in order to lose its giant component.

Fig. \ref{f3}(c) shows the results for the third network, Gnutella peer-to-peer network \cite{gnutella}. This is a network of Gnutella hosts from 2002. The nodes represent Gnutella hosts, and the directed edges represent connections between them. Again, the network is treated as undirected ignoring the direction of edges. For this network, $Commn$ and degree centrality work better, but performance of other centrality measures is also close.

Results of various centrality measures on the fourth network, World Wide Web \cite{www} are shown in Fig. \ref{f3}(d). In this network, each node represents a webpage and a directed link between the nodes shows a hyperlink from a webpage to another. Direction of the links have been ignored. For this network, $Commn$ centrality outperforms other centrality measures.

Hence, the proposed centrality measure performs at least as good as the global degree and in some cases, better than betweenness  and community-based mod centralities with less information about the network structure and in lesser time.

To quantify the dynamical differences in the epidemic spreading process when a fraction of nodes is immunized by using the proposed immunization strategy, time evolution plots are generated. Time evolution plots of, respectively, the fraction of the total nodes in \% of susceptable, infected and recovered population in the network with immunization and without immunization, are displayed in Fig. \ref{f21}. $ \lambda=0.5$ \& $ \sigma=0.1 $ and $ \lambda = 1 $ \& $ \sigma=0.1 $ is considered for the time evolution plots to understand the dynamical differences.

The experimental results reveal that, the proposed $Commn$ centrality is effective in identifying the influential nodes to be selected for immunization to prevent or mitigate the epidemic spreading. The proposed measure is as effective as the global degree and betweenness centrality based methods, but do not require any information about the global structure of the network. Community-based centrality and Mod centrality also works as well but they have certain drawbacks as pointed out earlier. We have proposed an alternative centrality measure which can be employed to find out the influential nodes in networks with community structure. The investigated strategy also performs better than the stochastic strategies, Acquaintance and \textit{CBF} strategies, with only the information about the local (community-wise) structure of the network. This suggests that the local information is sufficient to design an immunization strategy.

\section{Conclusions} \label{s6}
There is no enough information available about the global structure of the underlying relevant contact network in order to control the epidemic spreading. Therefore, efficient immunization strategies are required that can work with the information available at the community level. Results of our investigation, on a realistic synthetic benchmark as well as real-world social networks, show that the community structure plays a major role in the epidemic dynamics. It is observed that the proposed $Commn$ centrality based immunization strategy is effective in controlling the epidemic spreading. It works as well as the global centrality measures (degree and betweenness), without any knowledge of the global network structure.  This centrality measure defined at the community level is a good approximation of global centrality measures. The in-degree part of the $Commn$ centrality of a node represents its degree and betweenness centralities, relative to its own community, whereas, the out-degree part represents these global centralities relative to the other communities of the network. This strategy is also sensitive to the community structure of the network and hence automatically selects the influential nodes which have got a good balance of inner and outer connections to their community. Hence, this strategy is quite efficient as it generalizes to the networks with varying strength of community structure. Furthermore, unlike global centrality measures, the proposed measures do not require any knowledge of the global network topology and thus can be easily and quickly computed. It also overcomes the drawbacks of the Mod strategy also proposed for networks with community structure. Furthermore, it performs better than the stochastic strategies, Acquaintance and \textit{CBF}. Finally, the main lesson of this work is that exploiting the local information on the network topology can be very effective in order to design  efficient immunization strategies that can be used in large scale networks. These preliminary results pave the way for more investigations on alternative community topological measures.

\section*{Acknowledgement}
The authors thank Mr. Upendra Singh (B.Tech., MANIT Bhopal, India) for implementing the SIR model of epidemics, used in this paper.
\section*{References}
\bibliographystyle{elsarticle-num}
\bibliography{arxiv}

\end{document}